\documentclass[showpacs,aps,physrev,superscriptaddress, twocolumn, nofootinbib]{revtex4-2}
\usepackage{graphicx}
\usepackage{amsthm,amssymb,amsmath,braket,mathdots}
\usepackage[dvipsnames]{xcolor} 
\usepackage{float}
\definecolor{lightgreen}{RGB}{0, 200, 0}
\usepackage[colorlinks=true, urlcolor=blue, linkcolor=cyan, citecolor=lightgreen]{hyperref} 

\begin{document}

\title{Coherence and Transients in Nonlocally Coupled Dissipative Kicked Rotors}
\author{Jin Yan}
\email{jin.yan@wias-berlin.de}
\affiliation{\small Weierstrass Institute for Applied Analysis and Stochastics, Anton-Wilhelm-Amo-Stra{\ss}e 39, 10117 Berlin, Germany}

\begin{abstract}
The dynamics of nonlocally coupled dissipative kicked rotors is rich and complex. In this study, we consider a network of rotors where each interacts equally with a certain range of its neighbors. We focus on the influence of the coupling strength and the coupling range, and show both analytically and numerically the critical transitions in the phase diagram, which include bifurcations of simple spatiotemporal patterns and changes in basin sizes of coherent states with different wavenumbers. We highlight that this diagram is fundamentally different from those found in other coupled systems such as in coupled logistic maps or Lorenz systems. Finally, we show an interesting domain-wall phenomenon in the coupled chaotic rotors, where a super-long transient interface state (partially regular and partially chaotic) is observed and can persist exponentially long as the coupling range increases up to a critical threshold. 
\end{abstract}

\maketitle

\section{\label{sec:intro}Introduction}
The emergence of synchronized and correlated dynamics among coupled elements are fundamental phenomena observed in a wide range of complex systems, from biological neural networks to laser arrays and chemical oscillators. 
These behaviors typically emerge from the interplay between interactions and intrinsic properties \cite{pikovsky2001universal}. 
Beyond full synchronization, a variety of structured yet non-chaotic states, such as phase-locking \cite{donato2007phase}, metastability \cite{shanahan2010metastable, tognoli2014metastable}, cluster formation \cite{pecora2014cluster} and partial synchronization \cite{abrams2004chimera, hart2016experimental}, illustrate how systems can transiently coordinate while retaining diverse localized activities. 

In many real-world settings, interactions are neither purely nearest-neighbor (local) nor all-to-all (global, where every element interacts equally with all others), but extending over finite spatial ranges, leading to nonlocal couplings. Such couplings give rise to rich spatiotemporal patterns that are often absent in local or global cases. One of the most well-known examples is chimera states, characterized by the coexistence of coherent and incoherent domains \cite{panaggio2015chimera, scholl2016synchronization}. The interplay between the coupling range and strength makes the transition between order and disorder more intricate and complex \cite{tonjes2010synchronization, omelchenko2011loss}. 

However, the analysis of nonlocal coupling presents significant theoretical challenges, as it eludes the simplifying symmetries of all-to-all coupling via mean-field approaches (e.g., the Kuramoto model reduces to a single order parameter), and the spatial regularity of nearest-neighbor coupling via reaction-diffusion PDEs or discrete Laplacians. 
Nonlocal systems, with their distance-dependent interactions -- require more sophisticated approaches. 
The seminal work \cite{omelchenko2011loss} demonstrated this through coherence-incoherence transitions in nonlocally coupled logistic maps, and confirmed that such a transition is universal among systems with diverse local dynamics. 
Inspired by this, we present an analytical and numerical study of various transitions in a nonlocally coupled system with a two-dimensional local map, specifically the dissipative kicked rotor \cite{zaslavsky1978simplest, russomanno2023spatiotemporally, yan2025bifurcations}, which has many applications in optics, engineering and quantum physics \cite{chai2018survival, groiseau2019spontaneous, chai2020enhancing, andersen2022classical}. 
Interestingly, despite the nonlocal coupling, linear stability analysis effectively captures the dynamics of the coupled system beyond the linearization regime; the resulting phase diagram exhibits qualitatively different structure compared to previously studied systems \cite{omelchenko2011loss}. 
Analyzing bifurcations of simple states reveals transition points between coherent states of different spatial periodicity, and between coherent and incoherent states. 
These will be illustrated in detail. 

The second part of our work investigates super-long transients in systems where the local map has both non-chaotic and chaotic attractors. This regime is generic and intermediate between regular and fully chaotic behavior. When initialized with a cluster-like configuration, the system eventually evolves toward a fully chaotic state; however, the lifetime of a cluster-like state can be exponentially long depending on the coupling range. Remarkably, beyond a critical point, a sudden transition occurs, causing the transient lifetime to collapse by several orders of magnitude. 
The significance of these findings lies in the observation that cluster-like states are often desirable in real-world applications. Such states can support structured behavior or partial predictability and thus useful in neuroscience, power grids and information processing \cite{davidsen2024introduction, ercsey2011optimization, boccaletti2000control, scholl2008handbook}. Alternatively, in other contexts, a rapid onset of full chaos may be desirable \cite{capeans2017partially, boccaletti2000control, scholl2008handbook, wang2016transient}. By tuning the coupling parameters, it becomes possible to sustain such states for a controllably long duration, offering a practical mechanism to delay the onset of full chaos, or to reduce transient time by adding more connections to the system. 

The paper is organized as follows. In Sec.\ref{sec:model} we introduce the model and parameters. In Sec.\ref{sec:pattern} we study analytically the transition from a homogeneous state to a temporal period-$2$ state with a characteristic wavenumber, and identify the phase transitions under variations of the coupling length and strength. These are supported by numerical simulations. In Sec.\ref{sec:interface} we explore transient behavior when local rotors have a dominant chaotic attractor. We give a heuristic understanding of the super-long transient time.

\section{\label{sec:model}Model}
We consider a chain of $N$ nonlocally coupled dissipative kicked rotors with periodic boundary conditions, described by 
\begin{equation}
\begin{split}
p_j(t+1) =& \gamma p_j(t) - K_0\sin \theta_j(t) \\
&+ \frac{K}{2P_c}\sum_{k = j-P_c}^{j+P_c}\sin (\theta_k(t) - \theta_j(t)) \\
\theta_j(t+1) =& \theta_j(t) + p_j(t+1) \quad (\text{mod } 2\pi)
\end{split}
\label{eq:system}
\end{equation}
where $(p_j(t), \theta_j(t)) \in (-\infty, +\infty)\times [0, 2\pi)$ denote the momentum and angle of the rotor $j \in \{1, 2, ... N\}$ at time $t \in \mathbb{N}_0$, $\gamma \in (0, 1)$ is the dissipation coefficient, $K_0 > 0$ is the local nonlinearity parameter, $K>0$ is the coupling strength, and $P_c \in \left\{ 1, 2, .., \frac{N}{2}\right\}$ (assuming $N$ is even) is the coupling length, denoting the number of neighbors in each direction that are coupled to the rotor $j$. 
The momenta are set to be unbounded and non-periodic for physical relevance, though this choice does not affect the critical transitions discussed in this work.

The local map describes a dissipative kicked rotor and for a fixed $\gamma$, multiple bifurcations occur as $K_0$ is varied \cite{yan2025bifurcations}: the zero state $(p, \theta) = (0, 0)$ is the only attractor when $K_0$ is very small, and undergoes a period-doubling (PD) bifurcation at $K_0^{\text{PD}} = 2(1+\gamma)$. Due to periodicity of $\theta$ and non-periodicity of $p$, a series of attractors is created by fold bifurcations at discrete momentum levels $p = 2n\pi$, $n \in \mathbb{Z}$. These attractors further bifurcate into period-$2$ orbits and finally into chaos. The bifurcation scenarios for various values of $\gamma$ are illustrated in Fig.\ref{fig:srbifur} in Appendix \ref{app:gamma}, see also \cite{yan2025bifurcations}. Smaller $\gamma$ indicates stronger damping and results in more restricted bifurcations. 
At $\gamma=0$, the map reduces to the one-dimensional Arnold circle map, $\theta(t+1) = \theta(t) - K_0\sin \theta(t)$ (mod $2\pi$), which is a fundamental model of phase locking. In contrast, it becomes the Chirikov standard map at $\gamma=1$, a classical low-dimensional example of Hamiltonian chaos where KAM tori coexist with the chaotic sea. 
The qualitative behavior of the local map is unaffected by $\gamma \in (0, 1)$ and, as demonstrated in Appendix \ref{app:gamma}, the coupled system exhibits similar critical transitions across different values of the local parameters $\gamma$ and $K_0$.

In this paper, we fix $\gamma = 0.8$ without loss of generality \cite{yan2025bifurcations}, use $K_0=2$ (local map is non-chaotic and has three fixed points as the only attractors) for Sec.\ref{sec:pattern}, and use $K_0=6.6$ (local map has coexisting non-chaotic and chaotic attractors) for Sec.\ref{sec:interface}. 
Our focus is on the dynamics arising from the interplay between the coupling strength $K$ and the coupling length $P_c$.

\section{\label{sec:pattern}Spatial and Temporal Patterns}
The system \eqref{eq:system} possesses a large number of attractors due to multistability of the local map \cite{yan2025bifurcations} and the large system size $N$. Depending on the initial conditions, the rotors self-organize into temporally periodic patterns with or without a distinguishable spatial period. Consequently, for any given values of the coupling parameters $(P_c, K)$, multiple attractors can coexist. 
However, we notice that for a specific parameter region, the dominant attractors can be characterized by low temporal and spatial periods. 

In this section, we focus on the following three issues. The first one is the transition from the stationary, fully synchronized state to a coherent state of temporal period-$2$, which can be explained analytically through linear stability analysis. A wavenumber will be defined to characterize the spatial periodicity. 
The second issue is to study coherent regions on the parameter $(P_c, K)$-plane, particularly the regions with low spatial and temporal periods. The transition curves can be obtained from stability analysis of the temporal period-$2$ states. 
The third issue concerns the coexistence of the temporal period-$2$ states with different wavenumbers, and we numerically examine how their relative basin sizes vary with the coupling length $P_c$. 

Throughout this section, purely random initial conditions are used for numerical simulations: $(p_j(0), \theta_j(0)) \in \text{Uni}[-p_0, p_0]\times \text{Uni}[0, 2\pi)$, where $p_0>2\pi$ is constant (the specific value of $p_0$ does not affect the qualitative results).

\subsection{\label{subsec:sync}Transition from zero homogeneous states} 
\begin{figure*}
\centering
\includegraphics[width=0.95\linewidth]{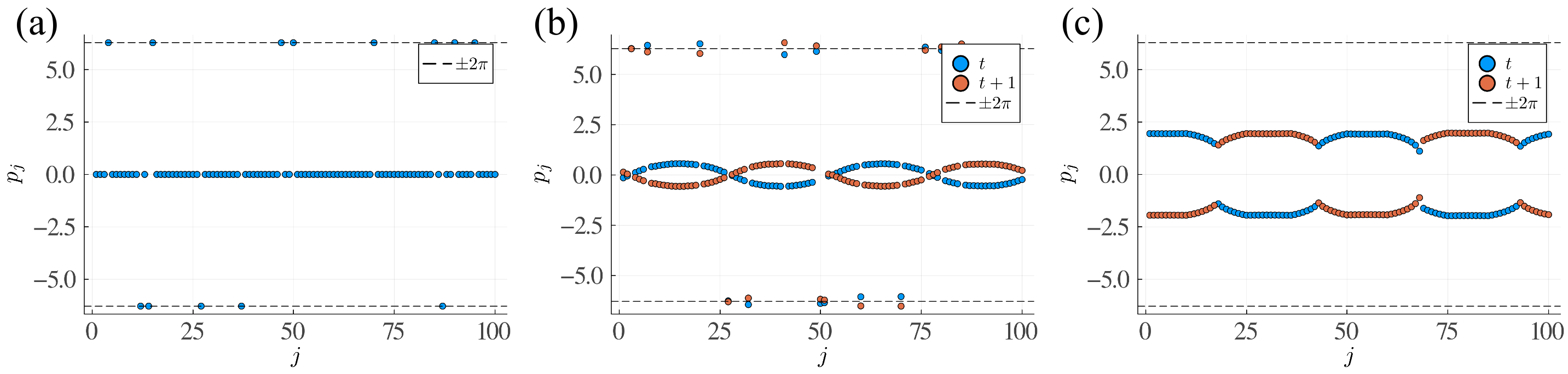}
\caption{Snapshots of momenta for (a) $K=1.3$, (b) $K=1.4$ and (c) $K=3$. Other parameters: $\gamma=0.8$, $K_0=2$, $P_c=32$, $N=100$ and random initial conditions $(p_j(0), \theta_j(0)) \in \text{Uni}[-35, 35]\times \text{Uni}[0, 2\pi)$, $j = 1, 2, ..., N$. The corresponding angles behave similarly.}
\label{fig:sync}
\end{figure*}

The full phase space of the system \eqref{eq:system} is a product of $N$ infinite cylinders $(p_j, \theta_j) \in (-\infty, +\infty) \times [0, 2\pi)$, $\forall j =1, 2, ..., N$. 
For $K_0$ and $K$ both small, trajectories initially close to $\boldsymbol{p} := (p_1, p_2, ..., p_N) = \boldsymbol{0}$ will converge to the {\it zero synchronized state} $(\boldsymbol{p}, \boldsymbol{\theta}) = (\boldsymbol{0}, \boldsymbol{0})$ and it appears as the only attractor. 
However, due to multistability of the local map, more attractors can be observed when we consider longer truncated cylinders (i.e., $\boldsymbol{p}$ further away from $\boldsymbol{0}$), and the notion of the zero synchronized state is extended to a state in which most rotors are homogeneous but with a few exceptions; these exceptional rotors locate randomly depending on the initial condition, and stay at the nonzero fixed points of the local map. We refer such a state a {\it zero homogeneous state}. An example is shown in Fig.\ref{fig:sync}(a): for $K_0=2$ the local map has three fixed points $p = 0, \pm 2\pi$, where $p=0$ has the largest basin and $p=\pm 2\pi$ have equal basin sizes \cite{yan2025bifurcations}. 

This kind of zero homogeneous states undergoes a transition to a temporal period-$2$ state with a spatial periodicity, an example is shown in Fig.\ref{fig:sync}(b). Further increasing the coupling strength $K$ reduces the number of exceptional rotors, leading to a more coherent pattern, cf. Fig.\ref{fig:sync}(c). 

This period-doubling bifurcation can be understood analytically via linear stability analysis, giving the critical parameter values at which the Jacobian eigenvalues first cross the unit circle (details see Appendix \ref{app:linear}): 
\begin{equation}
\begin{split}
K &= [2(1+\gamma) - K_0]\frac{P_c}{P_c - S_{\min}}, \\
S_{\min} &:= \min_w \left[\csc\frac{w}{2}\sin\frac{P_cw}{2}\cos\frac{(P_c+1)w}{2}\right], 
\label{eq:stability}
\end{split}
\end{equation}
where $w = \frac{2\pi l}{N}$, $l=0, 1, ..., N-1$ (for periodic boundary conditions). 
Note that, in the case of nearest-neighbor coupling ($P_c=1$), $S_{\min}=-1$ at $w=\pi$, giving the critical parameter curve $K_0=-2K+2(1+\gamma)$, which is consistent with the result in \cite{yan2025bifurcations} (where $K\equiv 2J$); moreover, $w=\pi$ being the most unstable mode indicates that every two consecutive rotors form a wave packet, referred to as the {\it alternating state} in \cite{yan2025bifurcations}.

We can also infer the spatial period of the bifurcated state by the value of $w$ (thus $l$) at criticality. 
For example, for the patterns illustrated in Fig.\ref{fig:sync} ($N=100, P_c=32$), we have $\tilde{S}_{\min} = \min_w \left[ \csc\frac{w}{2}\sin\frac{3w}{2}\cos\frac{33w}{2} \right] \approx -7.56572$ at $\tilde{w} \approx 0.13827$, by $w = \frac{2\pi l}{N}$, we have $\tilde{l} \approx 2.2$. Since the modes $w$ (and $l$) are discrete, we have denoted the continuous values by tilde. The closest integer is $l=2$ and the corresponding $S_{\min} \approx -6.94219$. The bifurcation point given by Eq.\eqref{eq:stability} is therefore $K \approx 1.31$. We refer the spatial period $l = 2$ as the {\it wavenumber}. 

\begin{figure*}
\centering
\includegraphics[width=0.95\linewidth]{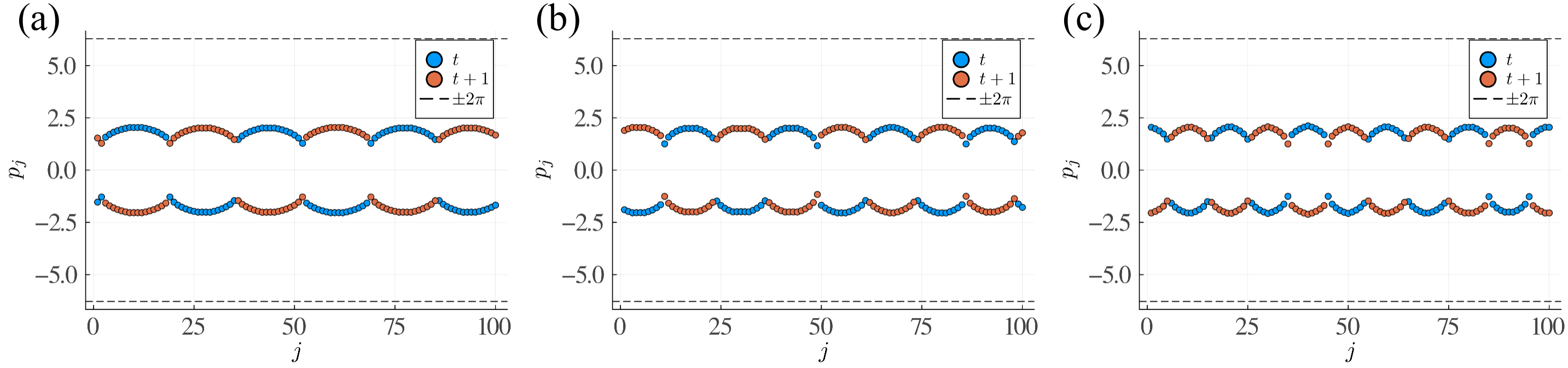}
\caption{Snapshots of momenta for (a) $P_c=23$, (b) $P_c=17$ and (c) $P_c=14$. Other parameters: $\gamma=0.8$, $K_0=2$, $K=3$, $N=100$ and random initial conditions $(p_j(0), \theta_j(0)) \in \text{Uni}[-35, 35]\times \text{Uni}[0, 2\pi)$, $j = 1, 2, ..., N$. The corresponding angles behave similarly.}
\label{fig:patterns}
\end{figure*}

For the system of size $N=100$ to form a wavenumber $l=1$, one would require $w=\frac{2\pi}{100}$ and the critical $P_c$ should thus be given by $S_{\min} = S\left( \frac{2\pi}{100}\right)$, which could be achieved only when $P_c=71 > \frac{N}{2}$. Thus, the lowest realizable wavenumber for the $N=100$ system is $l=2$. However, higher wavenumbers are possible: for example, an $l=3$ pattern emerges at $P_c=23$, $l=4$ at $P_c=17$, $l=5$ at $P_c=14$, and so on. Patterns with these wavenumbers are illustrated in Fig.\ref{fig:patterns}.

We note here that the wavenumber for the bifurcated state depends on the system size $N$. For a finite-size system, the eigenvalues $\lambda^-(l)$ are coarse-grained from the continuous ones $\lambda^-(\tilde{l})$, where, a coarser discretization (i.e., smaller $N$) would result in a less number of unstable eigenmodes. 
In a continuum limit ($N\to \infty$), the onset of instability occurs earlier than in any finite-size system. 
It implies that, while a smaller system realizes only one spatial pattern, therefore attracting all initial conditions in the phase space, a larger system can display multistability and realizes spatial patterns with different basin sizes. 
This will be discussed in the following subsection.

\subsection{\label{subsec:coherent}Coherent regions and their basin sizes} 
\cite{omelchenko2011loss, omelchenko2012transition} illustrates a generic phase diagram of spatiotemporal patterns in coupled systems, where a series of decreasing-sized tongue regions highlights the coherent patterns of an increasing wavenumber and (temporal) period-doubling within each tongue. The models studied therein are coupled logistic maps, Lorenz and R\"ossler systems, and parameters are chosen such that the local dynamics is chaotic. Similar diagrams are also observed in other systems such as Stuart-Landau oscillators \cite{zakharova2016amplitude} and coupled Chebyshev maps (Appendix \ref{app:cheby}). Here, with non-chaotic dissipative kicked rotor maps, we present that the phase diagram is qualitatively different from those in the above mentioned systems. 
Specifically, (i) there is no blowout bifurcation to full synchronization in the regime of large coupling length $P_c$ and large coupling strength $K$, (ii) there is no tongue-shaped region exhibiting a period-doubling cascade, and (iii) the coherence-incoherence transition does not appear to transit through partially synchronized (or chimera) states, or at least they are not dominant for any parameter values. 

\begin{figure*}
\centering
\includegraphics[width=0.65\linewidth]{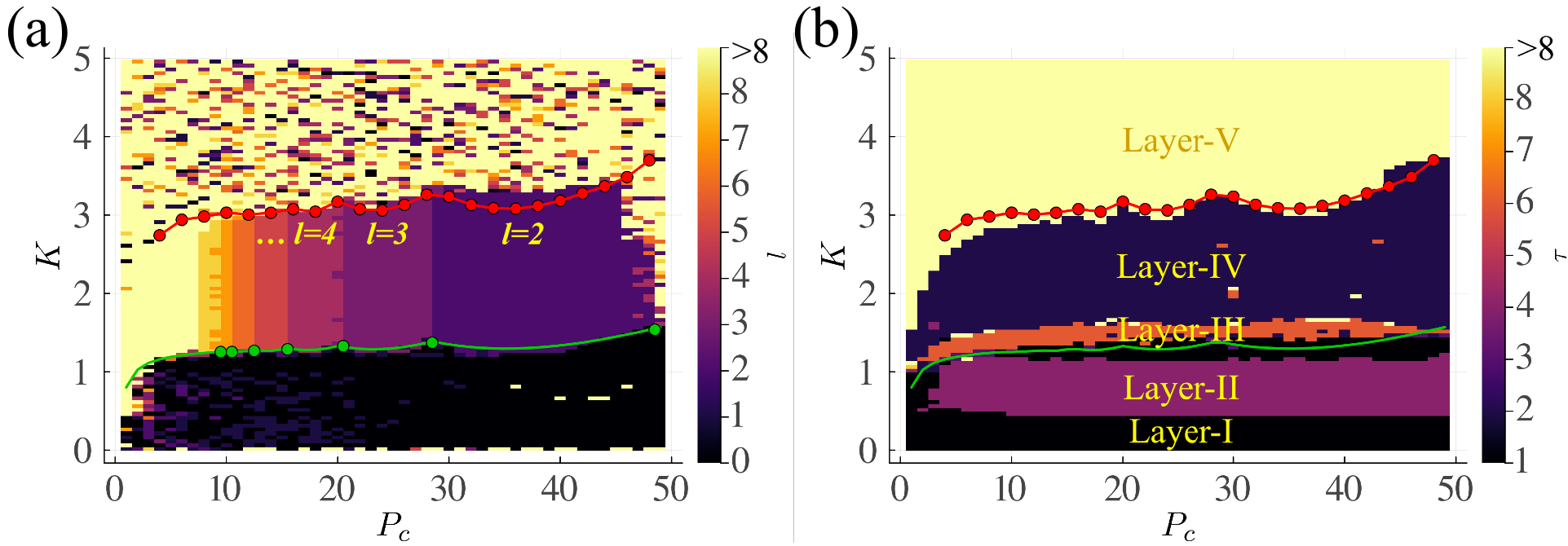}
\caption{Regions of typical (a) wavenumbers $l \leq 8$ and (b) temporal periods $\tau \leq 8$ on the parameter $(P_c, K)$-plane, with $P_c \in \{ 1, 2, ..., \frac{N}{2}-1\}$ and $K \in [0.01, 5]$ for $100$ equidistant values. Each pixel color is decided by the mode over $20$ trajectories. The green curve is obtained from linear stability analysis of the zero synchronized state and the green dots are the corresponding wavenumber changes. The red dots mark the instability of the temporal period-$2$ states, obtained by numerical bifurcation analysis. Other parameters: $\gamma=0.8$, $K_0=2$, $N=100$, and random initial conditions $(p_j(0), \theta_j(0)) \in \text{Uni}[-35, 35]\times \text{Uni}[0, 2\pi)$, $j = 1, 2, ..., N$. Larger-size systems show similar phase transitions, cf. Appendix \ref{app:large}.}
\label{fig:diagrams}
\end{figure*}

We plot the diagrams in Fig.\ref{fig:diagrams} for typical wavenumbers $l$ and temporal periods $\tau$ separately, where $l$ is determined by a numerical Fourier transform and $\tau$ by periodicity of the time series (up to numerical precision). 

We describe the different regions by the following five layers on the $(P_c, K)$-plane, and refer them in Fig.\ref{fig:diagrams}(b). The layers can be roughly characterized by the range of the coupling strength $K$: 
\begin{itemize}
\item Layer-I ($K\lesssim 0.4$): when the coupling is very weak, the zero homogeneous states ($l=0$, $\tau=1$) are dominant. This is well-understood from the uncoupled system that the zero fixed point has the largest basin compared to the other fixed points; 

\item Layer-II ($0.4\lesssim K\lesssim 1.1$): a type of temporal period-$4$ states with irregular spatial patterns dominates and coexists with the zero homogeneous states; note that it cannot be a synchronized state (i.e., not a low-dimensional attractor) since otherwise the system would reduce to the local map, where no stable period-$4$ state exists for $K_0=2$; 

\item Layer-III ($1.1\lesssim K\lesssim 1.5$): this is a transition regime where a type of temporal period-$6$ states with irregular spatial patterns dominates, but it appears only for a narrow window of $K$; note that here many patterns coexist, the example of $\tau=2$ shown in Fig.\ref{fig:sync}(b) are stable but its basin is relatively small compared to the $\tau=6$ states; 

\item Layer-IV ($1.5\lesssim K\lesssim 3.0$): temporal period-$2$ states with well-defined wavenumbers $l$ are dominant, which is a result from the period-doubling bifurcation of the zero homogeneous states explained in Sec.\ref{subsec:sync}; as $P_c$ decreases, coherent patterns exhibit increasing wavenumbers $l$ over progressively narrower sub-intervals of $P_c$; $l$ appears independent of $K$, resulting in vertical strips rather than tongue-shaped regions as shown in \cite{omelchenko2011loss, omelchenko2012transition}. We refer to this layer as the {\it coherent region}; 

\item Layer-V ($K\gtrsim 3.0$): chaotic motion dominates where the temporal period  is large and the wavenumber appears random. This transition from a stable low-period state direct into chaos is observed in the locally coupled case \cite{yan2025bifurcations}, as well as in other systems \cite{kawabe1991intermittent, beck2024generalization, bennett1990stability}.
\end{itemize}

The critical transitions in the diagrams can be explained by stability changes of the typical states: the green curve obtained from Eq.\eqref{eq:stability} marks the period-doubling bifurcations of the zero homogeneous states, with the green dots in Fig.\ref{fig:diagrams}(a) indicating the wavenumber change, also predicted by Eq.\eqref{eq:stability}. 
The boundary of chaos is given by the loss of stability of the temporal period-$2$ states, see the red curve\footnote{Since the parameter $P_c$ is discrete, the numerical continuation method does not apply here. Instead, we obtain the bifurcation points $(P_c, K)$ discretely and interpolate the curve.} obtained by numerical bifurcation analysis implemented in BifurcationKit.jl in Julia \cite{veltz:hal-02902346}. 
Both curves agree excellently with the transitions observed in simulations. 

We emphasize that the diagrams show statistical behavior that only the most probable patterns are marked. 
In the coherent region (Layer-IV), states with different wavenumbers $l$ coexist especially near the boundaries between adjacent $l$ values. 
This can be characterized by the {\it relative basin size}, defined as the probability of reaching an attractor from a random initial condition. It is commonly used in many-body systems \cite{wiley2006size, zhang2024deeper} where the basins of attraction lie on a high-dimensional subspace in a highly nonlinear manner. 

\begin{figure}
\centering
\includegraphics[width=0.9\linewidth]{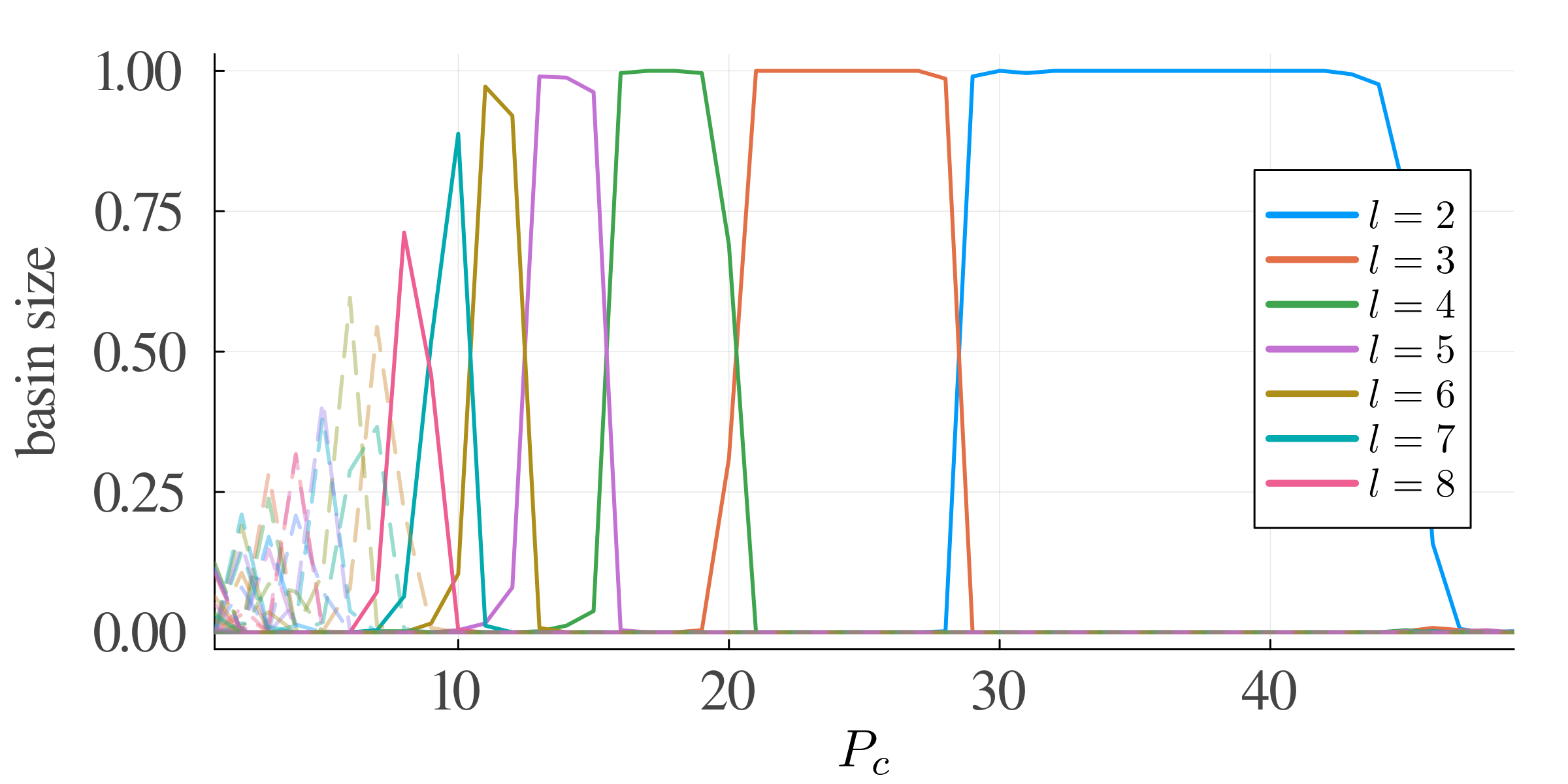}
\caption{Relative basin sizes for temporal period-$2$ states with wavenumbers $l$ as a function of the coupling strength $P_c$. The dashed lines denote $l > 8$. Other parameters: $\gamma=0.8$, $K_0=2$, $K=1.5$, and $N=100$. }
\label{fig:basins}
\end{figure}

As observed in the coherent region (Layer-IV) in Fig.\ref{fig:diagrams}, the wavenumber $l$ is independent of the coupling strength $K$, so we fix $K=1.5$ in this region. 
We simulate from a large number of random initial conditions\footnote{Note that since the whole phase space of the coupled system is a product of infinite cylinders, $M := ((-\infty, +\infty)\times [0, 2\pi))^N$, in practice we restrict to a subspace, namely, $([-p_0, p_0]\times [0, 2\pi))^N$ with $p_0=35$, so the basin sizes are conditioned on this subspace, where all the coherent states considered here are in this subspace.} and compute the relative basin sizes as $P_c$ varies. 
In Fig.\ref{fig:basins}, smaller wavenumbers showing in the legend correspond to those in the coherent region in the phase diagram Fig.\ref{fig:diagrams}(a); larger wavenumbers are dominant for smaller $P_c$ and are shown in dashed curves. 
As observed, for $P_c \geq 10$, a single wavenumber tends to dominate and can act as the unique attractor, with its relative basin size reaching $1$. In contrast, for smaller $P_c$, multiple wavenumbers $l$ have comparable basin sizes and coexist simultaneously. 

\section{\label{sec:interface}Super-long transient interface dynamics}
We have seen in the previous section rich dynamical patterns in the coupled dissipative kicked rotors where the local map is non-chaotic, yet chaos can emerge when the coupling strength $K$ is large enough. 

In this section, we examine the reverse scenario: when the local map is chaotic, the coupled system can still generate partially non-chaotic patterns, and depending on the coupling length $P_c$, these patterns can persist for an exponentially long time. 
With a large system size, transients become physically irrelevant and a fully chaotic state can never be achieved practically \cite{crutchfield1988attractors}. 

It has been shown that for large $K_0$, the single-rotor system has coexistence of chaos and tiny regular regions \cite{yan2025bifurcations}, which is also a common feature in Hamiltonian systems such as Lorentz gases \cite{klages2019normal}. We now consider initially a chain of $N$ uncoupled rotors consisting of half-chain of chaotic rotors and the other half regular. It creates a domain wall between the two middle rotors $j=\frac{N}{2}, \frac{N}{2}+1$, and the wall is stationary. Upon coupling, the domain wall becomes an interface, chaotic and regular motions can penetrate each other. Due to their relative basin sizes at the single-rotor level, the chaotic domain gradually encroaches the regular domain and eventually all rotors are chaotic. This has been observed in other coupled systems and has important implications in fluid dynamics \cite{keeler1986robust}. 

Our interesting observation is that, depending on the parameter values $(P_c, K)$, there is a sharp transition in the transient time before the system reaches a fully chaotic state. 

\begin{figure*}
\centering
\includegraphics[width=0.95\linewidth]{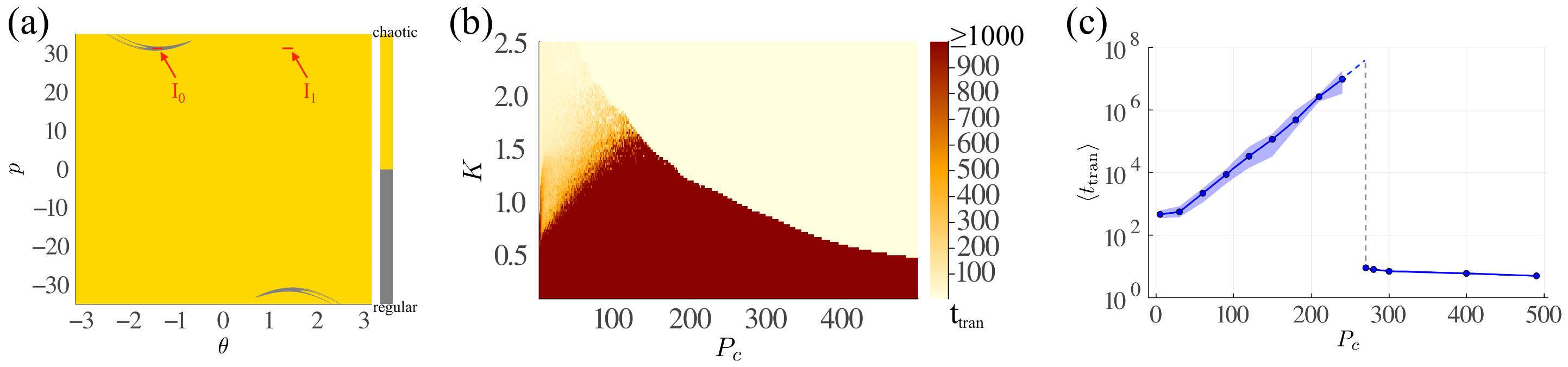}
\caption{Transient behavior for an $N=1000$ system: (a) basin structure of the single rotor at $\gamma=0.8$ and $K_0=6.6$, the two sub-intervals are $I_0 := [\theta^* -0.05, \theta^* +0.05]$ with $p_j(0) = p^*$, where $(p^*, \theta^*) = (10\pi, \arcsin \frac{-(1-\gamma)p^*}{K_0})$ is the fixed point of the single-rotor model and $I_1 := [-\theta^* -0.05, -\theta^* +0.05]$; $I_0$ ($I_1$) is a subset of the basin the regular (chaotic) attractor; (b) averaged transient time on the parameter $(P_c, K)$-plane and (c) for $K=1$ in a semi-log scale: simulation data are shown in blue dots with fluctuations in blue ribbons; the gray dashed line marks the sharp transition.}
\label{fig:ttran}
\end{figure*}

To illustrate this phenomenon, we choose $K_0=6.6$ so that the single rotor has coexistence of a regular and a chaotic attractor, whose basin is shown in gray and yellow respectively in Fig.\ref{fig:ttran}(a). For simplicity, initial points are prepared such that rotors $j=1, 2, ..., \frac{N}{2}$ are chosen randomly in a line segment $I_0$ inside the regular basin, and rotors $j=\frac{N}{2}+1, ..., N$ are chosen in its counterpart, $I_1$, in the chaotic basin (this particular choice does not influence the dynamics since the chaotic basin is mixing). 

Fig.\ref{fig:ttran}(b) shows the averaged transient time for a system of $N=1000$ rotors with different values of $(P_c, K)$. The iteration time $t=1000$ is sufficient to detect the transition, i.e., the boundary separating long (dark-red) and short transient (light-yellow) regimes. The averaged transient time at $K=1$ is illustrated in Fig.\ref{fig:ttran}(c): it increases exponentially with $P_c$ until the transition occurs. 

\begin{figure*}
\centering
\includegraphics[width=0.95\linewidth]{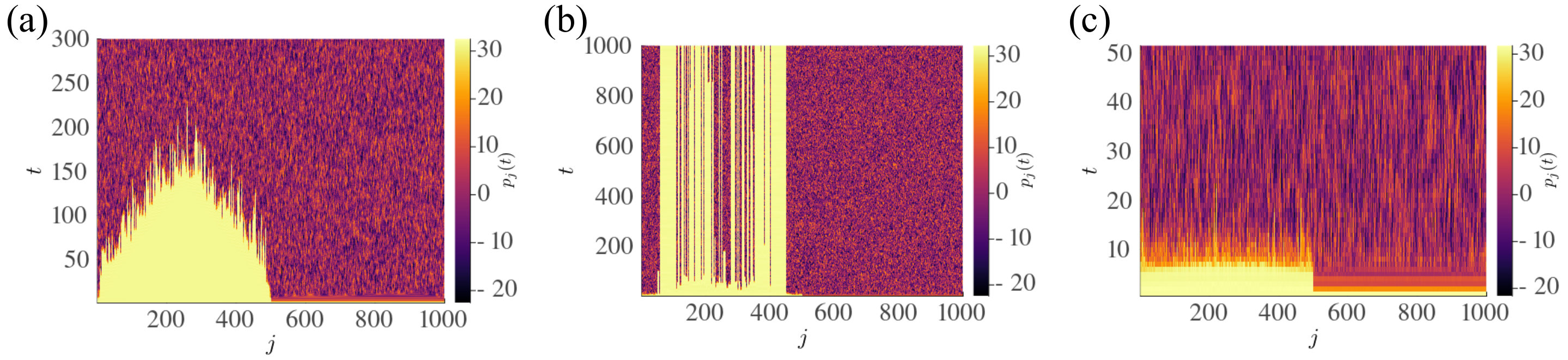}
\caption{Spatiotemporal evolutions of the rotor momenta $p_j(t)$ for $K=1$: (a) a short transient at $P_c=10$, (b) a long transient at $P_c=100$, and (c) a very short transient at $P_c = 300$. Other parameters are the same as in Fig.\ref{fig:ttran}(b).}
\label{fig:ttranEgs}
\end{figure*}

Fig.\ref{fig:ttranEgs} illustrates spatiotemporal evolutions of the $N=1000$ system for $P_c = 10, 100$ and $300$, showing three different regimes of the transient behavior: 
when the coupling length $P_c$ is very small (Fig.\ref{fig:ttranEgs}(a)), each rotor experiences localized interactions, and the domain wall propagates at a fixed speed in space, giving a short and predictable transient time. As $P_c$ increases, the interactions become less local and a competition between chaos (as forcing or reaction) and coupling (as diffusion) emerges. In this regime, most regular rotors sustain themselves for an exponentially long time, see Fig.\ref{fig:ttranEgs}(b). For very large $P_c$, the interactions are almost global, and the system effectively experiences a mean-field-like force, causing all rotors to rapidly ``synchronize" to chaotic motions. In this case, the partially non-chaotic state survives for only $8\sim 10$ iterations (Fig.\ref{fig:ttranEgs}(c)). 
However, identifying precisely this transition curve is nontrivial and remains an open question for future work. 

\section{Conclusion}
In this paper, we studied dynamics of a system of nonlocally coupled dissipative kicked rotors under variations of the coupling strength and the coupling length. 
We first analyzed the transition from a stationary homogeneous state to a coherent state of temporal period-$2$ via a period-doubling bifurcation. The spatial periodicity of coherent states are characterized by a wavenumber. As the coupling length decreases, the dominant wavenumber increases, which is also shown in the change of basin sizes. However, it is independent of the coupling strength in a certain range. 
These behaviors are summarized in the parameter space in Fig.\ref{fig:diagrams}, where the region of coherent states,  consisting of a sequence of reducing-sized strips labeling an increasing wavenumber, is bounded by the stability curves of the temporal period-$2$ states. 

In the last section, we explored numerically a super-long transient phenomenon when the single rotor has coexistence of regular and chaotic attractors. At intermediate coupling lengths, the lifetime of a partially regular and partially chaotic state grows exponentially. Beyond a critical point, the transient time abruptly collapses to just a few iterations, and the system rapidly transitions to full chaos.

Many interesting questions remain open. For example, we did not discuss the transitions among the Layer-I and -II, Layer-III and -IV in Fig.\ref{fig:diagrams}(b) as they are not the main focus of this paper, but it would be interesting to study them in detail.
Second, the instability of the temporal period-$2$ states -- marking the onset of chaos without developing a period-doubling (or multiplying) cascade -- reveals a fundamental difference from the behavior in \cite{omelchenko2011loss} (or in Appendix \ref{app:cheby}). If we compare the local map, the single rotor exhibits a period-doubling cascade to chaos \cite{russomanno2023spatiotemporally}, but the cascade occurs almost immediately and does not follow the Feigenbaum universality \cite{yan2025bifurcations}; while the logistic map (or Chebyshev maps in Appendix \ref{app:cheby}) follows the Feigenbaum universality. This may explain the difference. 
But the question remains: are both phase diagrams generic, and under what conditions? 
Moreover, while we have illustrated the relative basin sizes of the different wavenumber states, the geometry of their basins of attraction changes in the phase space remains unclear. 
For the super-long transient interface phenomenon, one could explore alternative initial cluster configurations, for example, setting half-chain at one fixed point and the other half at the other fixed point. Given the dominance of the chaotic basin, such configurations will eventually reach fully chaotic states, but the interface would evolve differently than presented here. Finally, we leave the analytical prediction of the sharp lifetime transition for future work. 

{\bf Acknowledgments} The author would like to thanks the two anonymous referees for their suggestions, which helped improve the quality and clarity of this manuscript.

\bibliography{refPRE}

\appendix
\onecolumngrid
\renewcommand{\thefigure}{A\arabic{figure}}
\setcounter{figure}{0}
\section{\label{app:gamma}Influence of the dissipation coefficient $\gamma$}
We illustrate here that the qualitative dynamics of the single-rotor model and the existence of the coherent region in the coupled system are not influenced by the value of the dissipation coefficient $\gamma \in (0, 1)$. 

\begin{figure}[H]
\centering
\includegraphics[width=0.95\linewidth]{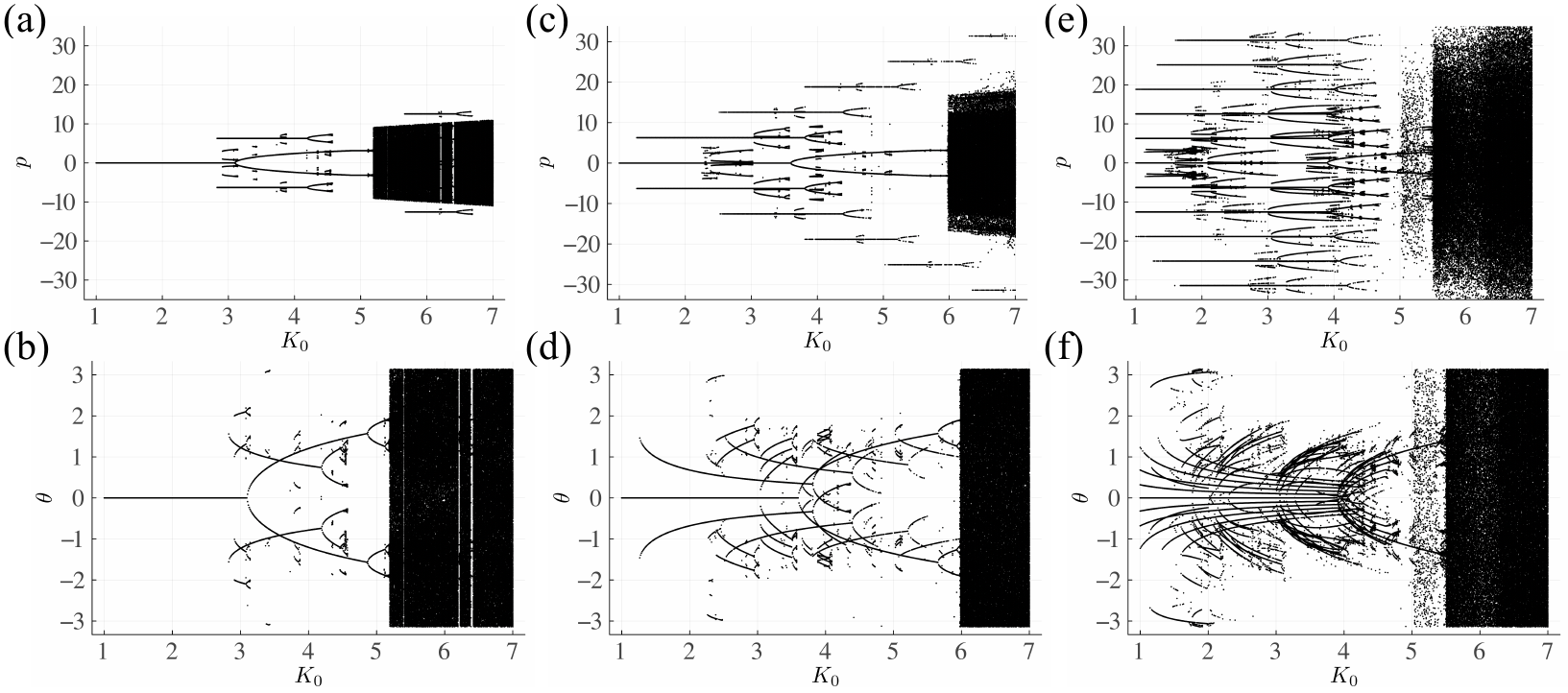}
\caption{Bifurcations in the single dissipative rotor with (a)-(b) $\gamma=0.55$, (c)-(d) $\gamma=0.8$ and (e)-(f) $\gamma=0.95$. The upper panel shows for momentum $p$ and the lower panel for angle $\theta$.}
\label{fig:srbifur}
\end{figure}

In Fig.\ref{fig:srbifur}, we plot, for three different values of $\gamma$, the bifurcations in the single-rotor model for the momentum $p$ (upper panel) and the angle $\theta$ (lower panel). Stronger dissipation (i.e., smaller $\gamma$) reduces the accessible momentum range and limits the number of possible bifurcating branches, but it does not alter the qualitative branching structure of period-doubling and fold bifurcations \cite{yan2025bifurcations}.

\begin{figure}[H]
\centering
\includegraphics[width=0.65\linewidth]{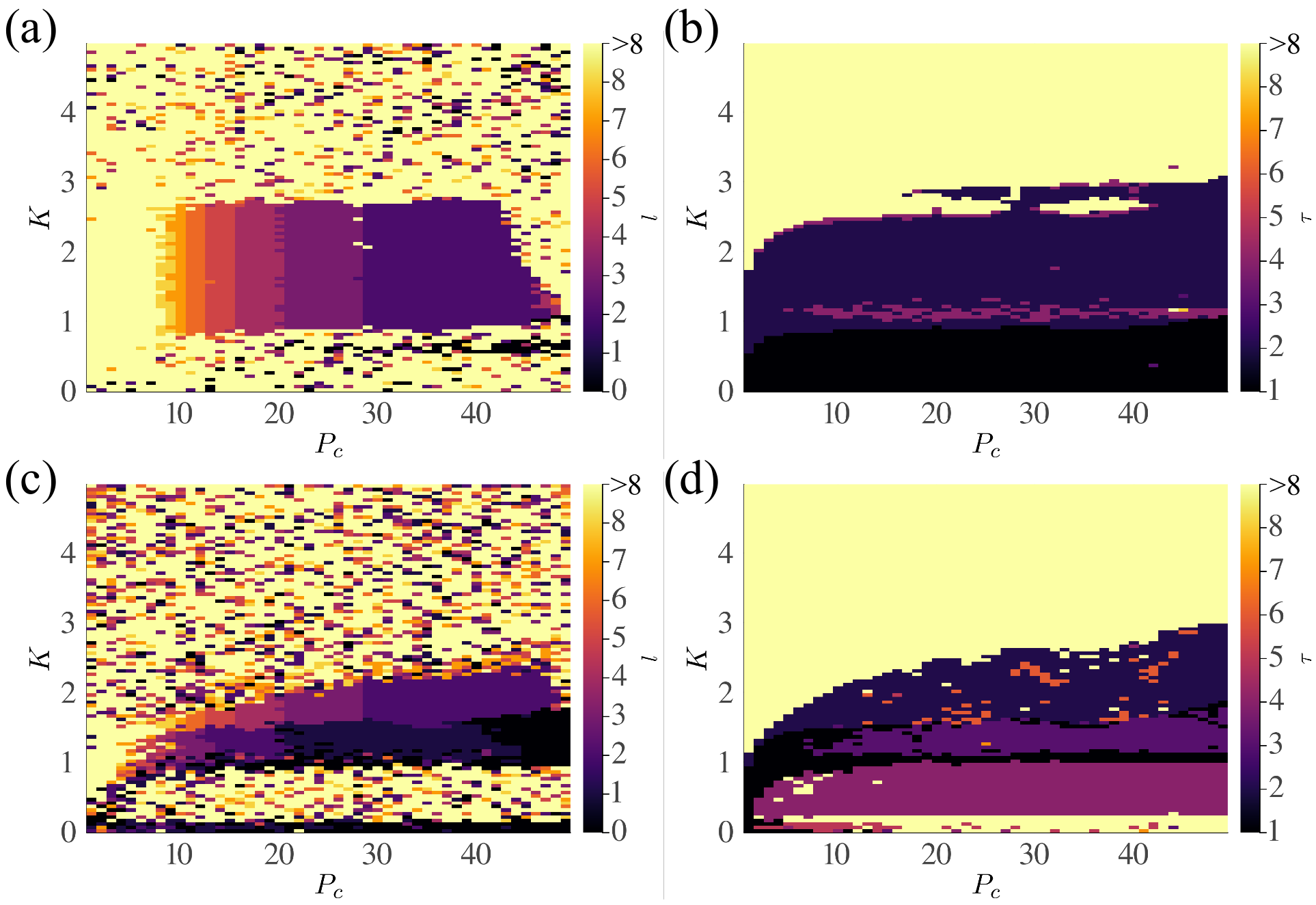}
\caption{Regions of typical wavenumbers $l\leq 8$ and temporal periods $\tau \leq 8$ on the parameter $(P_c, K)$-plane. Upper panel: $(\gamma, K_0) = (0.55, 2)$ and lower panel: $(\gamma, K_0) = (0.95, 2)$. Other numerical settings are as in Fig.\ref{fig:diagrams}.}
\label{fig:diagram-other-ga}
\end{figure}

On the other hand, the dynamics of the coupled system shown in Fig.\ref{fig:diagrams} is generic for various values of $\gamma$. 
Two examples are illustrated in Fig.\ref{fig:diagram-other-ga}: compared to $\gamma=0.8$ in the main text, the coherence region (Layer-IV in Fig.\ref{fig:diagrams}) is present for both stronger ($\gamma=0.55$) and weaker ($\gamma=0.95$) dissipation, which is characterized by temporal period $\tau=2$ and an increasing wavenumber $l$ over progressively narrower sub-intervals of the coupling length $P_c$. 
The only reason we use $\gamma=0.8$ together with $K_0=2$ in Sec.\ref{sec:pattern} is that, as shown in Figs.\ref{fig:srbifur}(c)-(d), the single rotor exhibits a relatively simple attractor structure, namely the three fixed points at $p=0, \pm 2\pi$. This simplicity carries over to the weakly coupled regime, allowing a clear demonstration of the critical transitions.

\section{\label{app:linear}Linear stability and behavior of the Jacobian eigenvalues}
\begin{figure}[H]
\centering
\includegraphics[width=0.95\linewidth]{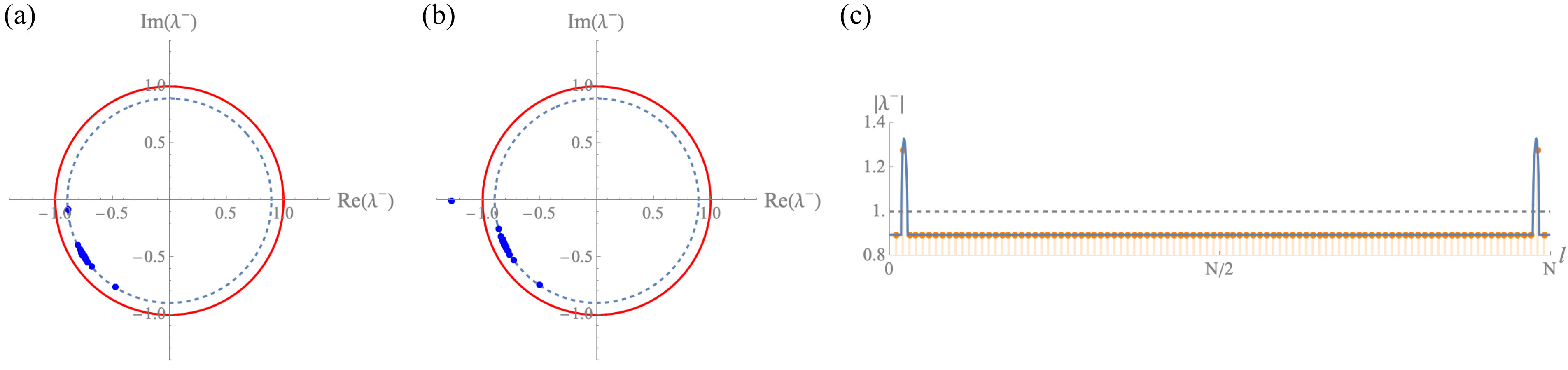}
\caption{Eigenvalues $\lambda^-$ before (a) $K=1.3$ and after (b)-(c) $K=1.4$ the period-doubling bifurcation; (c) shows $|\lambda^-|$ in continuous wavenumbers $\tilde{l}$ (blue) and in discrete $l$ (orange). Other parameter values: $\gamma=0.8$, $K_0=2$, $P_c=32$ and $N=100$.} 
\label{fig:evalues-K1pt3and1pt4}
\end{figure}

Under the assumption of slow angle variations, the equations of momenta in Eq.\eqref{eq:system} can be linearized as 
\begin{equation*}
p_j(t+1) = \gamma p_j(t) - K_0\theta_j(t) + \frac{K}{2P_c}\left[\sum_{k=j-P_c}^{j+P_c}\theta_k(t) - 2P_c\theta_j(t)\right]. 
\end{equation*}
By a Fourier transform $p_j(t) = \sum_wP_w(t)e^{iwj}$, $\theta_j(t) = \sum_w\Theta_w(t)e^{iwj}$, where $w=\frac{2\pi l}{N}$ and $l = 0, 1, ..., N-1$ (for periodic boundary conditions) we have, for each Fourier mode $w$, 
\begin{equation*}
P_w(t+1) = \gamma P_w(t) + \left[-(K_0+K) + \frac{K}{P_c}\sum_{k=1}^{P_c}\cos (wk) \right]\Theta_w(t).
\end{equation*}
The linearized equations of motion thus become
\begin{equation*}
\begin{split}
\begin{pmatrix}
P_w(t+1) \\
\Theta_w(t+1)
\end{pmatrix} &= \begin{pmatrix} 
\gamma & A \\
\gamma & 1+A
\end{pmatrix} \begin{pmatrix} 
P_w(t)\\
\Theta_w(t)
\end{pmatrix}, \\
A &:= -(K_0+K)+\frac{K}{P_c}\sum_{k=1}^{P_c}\cos(wk).
\end{split}
\end{equation*}
The characteristic equation $\lambda^2 - (\gamma+1+A)\lambda+\gamma=0$ gives the critical behavior at $|\lambda|=1$. We denote 
$$\lambda_w^{\pm} := \frac{1}{2}\left[(\gamma+1+A)\pm \sqrt{(\gamma+1+A)^2-4\gamma} \right].$$ 
The maximum of $|\lambda_w^+|$ occurs when $A = \max_w A = -K_0$ at $w=0$, and we have $\max_w|\lambda_w^+| = \frac{1}{2}[(\gamma + 1 - K_0) + \sqrt{(\gamma + 1 - K_0)^2 - 4\gamma}] < 1 + (\gamma - K_0) < 1$ for all parameter values under consideration ($\gamma \in (0, 1)$ and $K_0>1$). So the critical point is given by $\max_w|\lambda_w^-|=1$. Further, the observed period-doubling bifurcation indicates $\min_w\lambda_w^-=-1$, which gives $A=-2(1+\gamma)$. Since $\lambda_w^-(A)$ increases with $A$, this is achieved at $A = \min_w A$, i.e., $S(w) := \sum_{k=1}^{P_c}\cos(wk) = \csc\frac{w}{2}\sin\frac{P_cw}{2}\cos\frac{(P_c+1)w}{2}$ is minimized. Let us denote $S_{\min} := \min_w \left[\csc\frac{w}{2}\sin\frac{P_cw}{2}\cos\frac{(P_c+1)w}{2}\right]$, then the criticality occurs at 
\begin{equation*}
-2(1 + \gamma) = -(K_0 + K) + \frac{K}{P_c}S_{\min}, 
\end{equation*}
or equivalently, 
\begin{equation*}
K = [2(1+\gamma) - K_0]\frac{P_c}{P_c - S_{\min}}.
\end{equation*}
For the examples showed in Fig.\ref{fig:patterns}, we plot the eigenvalues $\lambda_w^-$ in Fig.\ref{fig:evalues-K1pt3and1pt4} before ($K=1.3$) and after ($K=1.4$) the period-doubling bifurcation. 
Fig.\ref{fig:evalues-K1pt3and1pt4}(c) illustrates that $l=2$ is the most unstable discrete eigenmode, corresponding to a spatial pattern of wavenumber $l=2$.

\begin{figure}[H]
\centering
\includegraphics[width=0.95\linewidth]{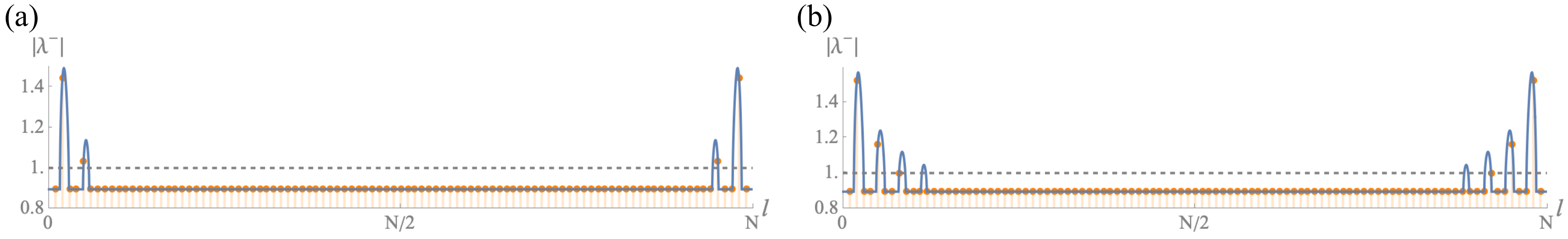}
\caption{$|\lambda^-|$ in continuous wavenumbers $\tilde{l}$ (blue) and in discrete $l$ (orange): (a) $K=1.48$ and (b) $K=1.52$. Other parameter values: $\gamma=0.8$, $K_0=2$, $P_c=32$ and $N=100$.} 
\label{fig:evaluesAbs2}
\end{figure}

Beyond the first instability, additional unstable modes emerge in the eigenspectrum. Fig.\ref{fig:evaluesAbs2} shows the cases after two and three successive crossings of the unit circle. 

\begin{figure}[H]
\centering
\includegraphics[width=0.45\linewidth]{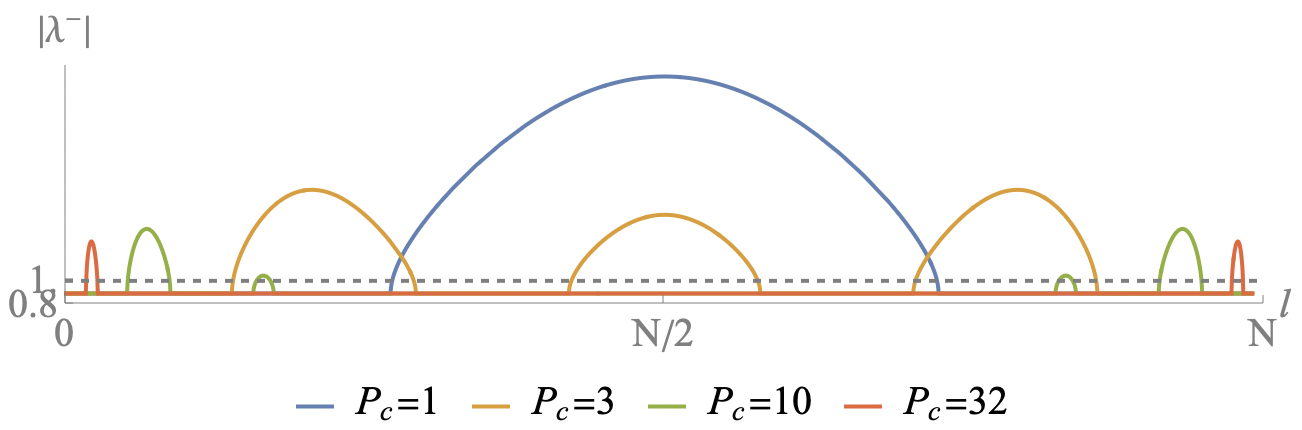}
\caption{$|\lambda^-|$ in continuous wavenumbers $\tilde{l}$ for different coupling lengths $P_c$. Other parameter values: $\gamma=0.8$, $K_0=2$, $K=1.4$ and $N=100$.} 
\label{fig:evaluesAbs3} 
\end{figure}

Fig.\ref{fig:evaluesAbs3} presents the influence of the coupling length $P_c$ on the unstable modes: for a fixed $K$, decreasing $P_c$ leads to larger unstable wavenumbers $l$. In particular, the nearest-neighbor coupling ($P_c=1$) yields $l=\frac{N}{2}$, corresponding to a spatially alternating pattern.

\section{\label{app:cheby}Coherent regions for coupled Chebyshev maps}
\begin{figure}[H]
\centering
\includegraphics[width=0.6\linewidth]{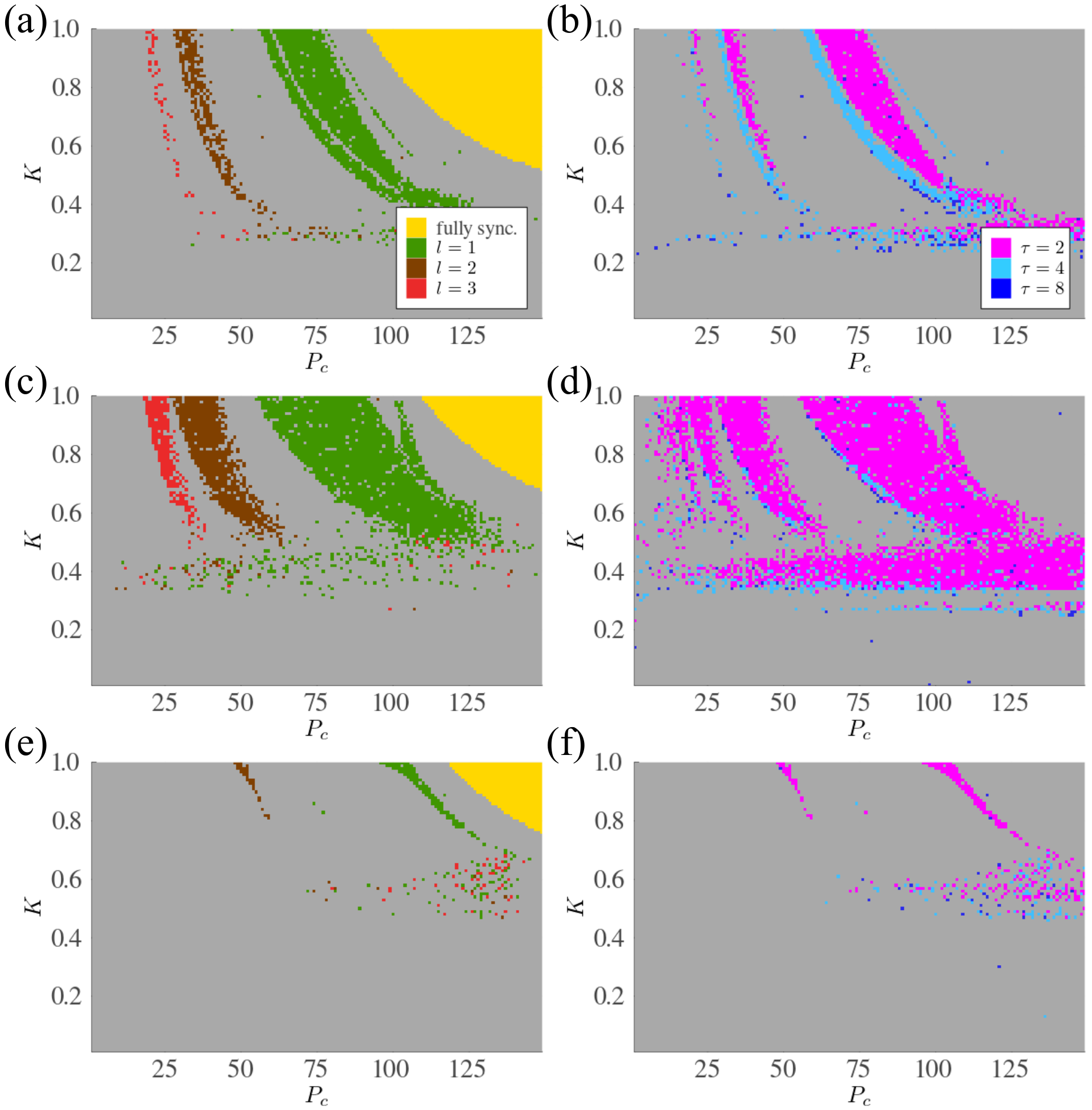}
\caption{Coherent regions for the system of nonlocally coupled Chebyshev maps $T_n$: (a)-(b) $n=2$, (c)-(d) $n=3$ and (e)-(f) $n=4$. The coupling scheme is the same as in Eq.\eqref{eq:system}. The left column shows wavenumbers $l = 1, 2, 3$ with temporal period $\tau \leq 8$ in green, brown and red, respectively, and the fully synchronized region in yellow; the right column shows temporal periods $\tau = 2, 4, 8$ in magenta, light blue and deep blue, respectively. All other patterns are colored in gray. Numerical settings: system size $N = 300$, $P_c \times K \in [1, 149]\times [0.01, 1]$ of resolution $149\times 100$.}
\label{fig:cheby}
\end{figure}

The $n$th-order ($n \geq 2$) Chebyshev maps \cite{yan2020distinguished} are defined via 
$T_n(x) = \cos(n\arccos(x))$, $x \in [-1, 1]$, 
which can be written as polynomials, for example, 
$T_2 (x) = 2x^2 - 1$, 
$T_3 (x) = 4x^3 - 3x$, 
$T_4 (x) = 8x^4 - 8x^2 + 1$, and so on.  
The $2$nd-order Chebyshev map is equivalent to the logistic map, except that it is upside-down and defined on the interval $[-1, 1]$. Therefore, the dynamics of the coupled $T_2$ system is expected to resemble that of the coupled chaotic logistic map \cite{omelchenko2012transition}, as confirmed in the first row of Fig.\ref{fig:cheby}. 
For higher orders $n$, the phase diagrams show similar patterns but the sizes of the tongue regions vary; see the last two rows in Fig.\ref{fig:cheby} for $n=3, 4$. 
In each left panel, the yellow region indicates a fully synchronized state, where the temporal dynamics follows the single Chebyshev map, which is chaotic. 

\section{\label{app:large}Coherent regions for coupled kicked rotors of larger system sizes} 
The coherent regions shown in Fig.\ref{fig:diagrams} are system-size independent. The plots below shows phase diagrams for $N=300$ and $1000$ systems. The diagram for the temporal period $\tau$ appears less layered when $N$ is large, due to an increasing number of exceptional rotors in the system. The overall pattern still follows the temporal periodicity $2$ in the coherent region. 

\begin{figure}[H]
\centering
\includegraphics[width=0.65\linewidth]{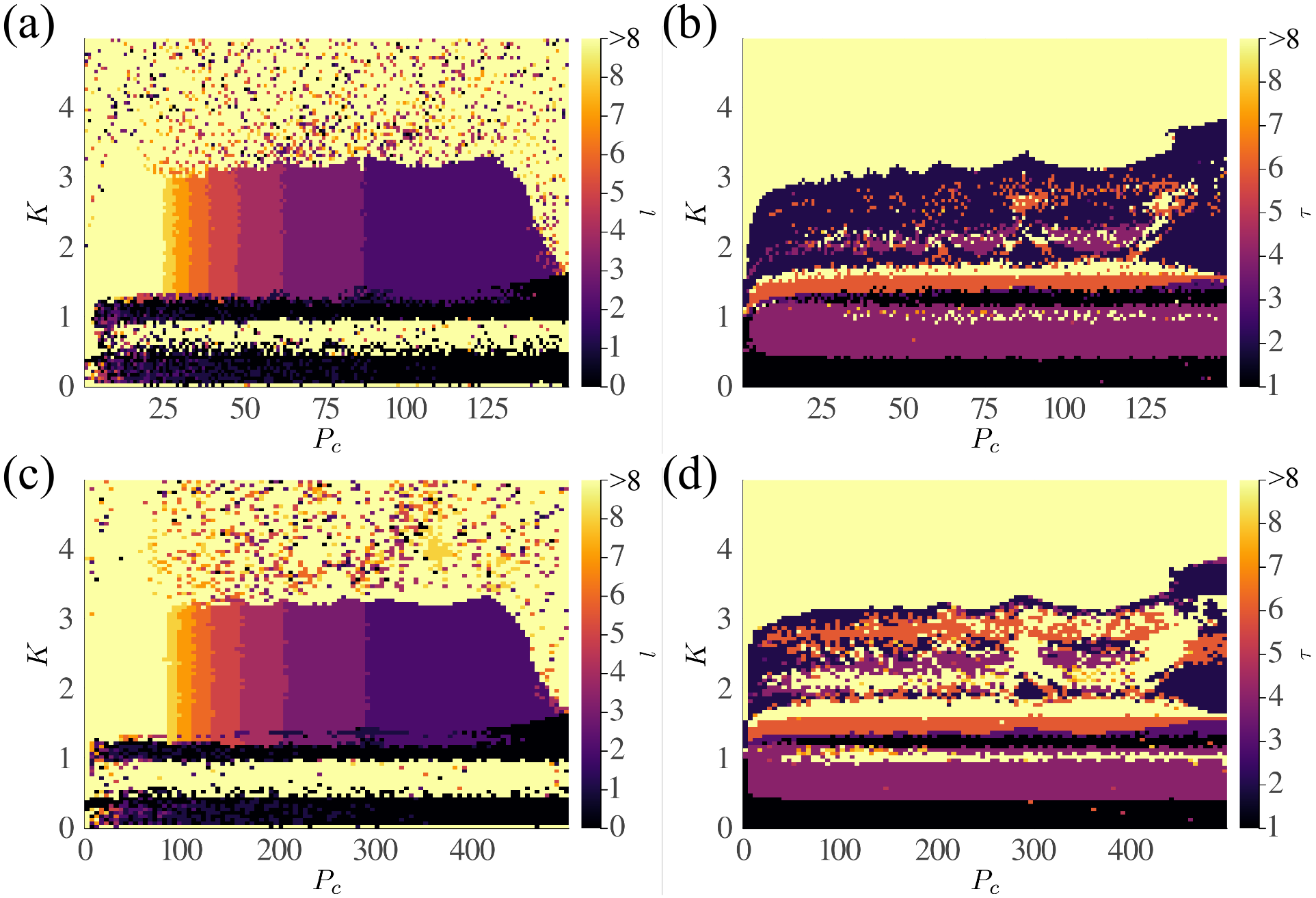}
\caption{Regions of typical wave numbers $l$ and temporal periods $\tau$ in the parameter $(P_c, K)$-plane for the system size (a)-(b) $N=300$ and (c)-(d) $1000$. $P_c \in \{ 1, ..., \frac{N}{2}-1\}$ and $K \in [0.01, 5]$ for $100$ equidistant values. Each pixel color is decided by the mode over $10$ trajectories. Other parameter values are as in Fig.\ref{fig:diagrams}. }
\end{figure}

\end{document}